\documentclass[twocolumn,aps]{revtex4}
\usepackage{graphicx}
\usepackage{amsmath}

\begin{document}

\title{Spontaneous Symmetry Breaking of Population between Two Dynamic Attractors
in a Driven Atomic Trap: Ising-class Phase Transition}
\author{Kihwan Kim$^{1,2}$, Myoung-Sun Heo$^{1,2}$, Ki-Hwan Lee$^{1,2}$, Kiyoub
Jang$^{1,2}$, Heung-Ryoul Noh$^{\dag}$, Doochul Kim$^{1}$, and
Wonho Jhe$^{1,2}$\footnote{Corresponding author: whjhe@snu.ac.kr}}

\affiliation{$^{1}$School of Physics and $^{2}$Center for Near-field Atom-photon Technology,\\
Seoul National University, Seoul 151-747, Korea \\
$^{\dag}$Department of Physics, Chonnam National University, Gwangju
500-757, Korea}

\begin{abstract}
We have observed spontaneous symmetry breaking of atomic
populations in the dynamic phase-space double-potential system,
which is produced in the parametrically driven magneto-optical
trap of atoms. We find that the system exhibits similar
characteristics of the Ising-class phase transition and the
critical value of the control parameter, which is the total atomic
number, can be calculated. In particular, the collective effect of
the laser shadow becomes dominant at large atomic number, which is
responsible for the population asymmetry of the dynamic two-state
system. This study may be useful for investigation of dynamic
phase transition and temporal behaviour of critical phenomena.


\end{abstract}

\date{\today}
\maketitle
%\section{Introduction}
The phenomena of symmetry breaking, widespread in nature with
examples from cosmology to biology, have been much studied
\cite{Demokritov, Lynden-Bell, Cambournac}. Recently in a
vibro-fluidized granular gas, spontaneous symmetry breaking (SSB)
of temperature and population between two compartments connected
by a hole was reported and understood in terms of the
density-dependent inelastic collision rates \cite{Eggers,
Montero}. Moreover there have been many works on
fluctuation-induced transitions in equilibrium \cite{Haenggi,
Simon, McCann} as well as far from equilibrium \cite{Luchinsky,
Penning, cnat2}. The double well structure of these systems is
very similar to the two compartments of the granular box, where
SSB was observed. In particular, we have recently studied the
atomic population transition between two dynamic phase-space
attractors available in the parametrically driven magneto-optical
trap (MOT) system \cite{cnat2}.

In this Letter, we report on experimental as well as theoretical
investigation of SSB of the atomic population between two dynamic
states of the driven MOT. We have found that the control parameter
for SSB is the total number of atoms in both states: The
population equality between the two equivalent states is broken
spontaneously above a critical number of atoms. This phenomenon
can be well understood as the Ising-class phase transition. We
have measured the critical number under various experimental
parameters and also observed the temporal evolution from symmetric
to asymmetric states above threshold. The SSB mechanism is
described qualitatively by considering two collective interactions
occurring at large atomic number, the shadow effect and the
reradiation effect \cite{Dalibard,Wieman,Bagnato,Wilkowski}. In
particular, the measured critical numbers are in good agreement
with the analytical and the simulational results.

\begin{figure}[t]
\hspace{-1.2cm}
\includegraphics[scale=0.4]{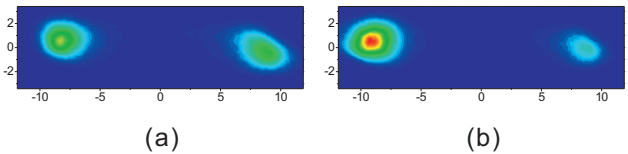}
\caption{{Snapshot images of atoms in two dynamic attractors (a)
before SSB of atomic population and (b) after SSB, taken by a
charge-coupled-device (CCD). The total number of atoms is (a)
6.1$\times 10^7$ and (b) 6.9$\times 10^7$, respectively. The
relative population difference in (b) is 0.63. Here $f$ = 96 Hz,
$h$ = 0.9, and the abscissas are in unit of mm. }}
\label{fig:1Image}
\end{figure}

%\section{Experimental Observations}
The experimental scheme is similar to those reported in previous
works on parametrically modulated MOT \cite{cnat1, cnat2}, where
we observed parametric excitation, limit cycles (dynamic
phase-space attractors), super-critical and sub-critical
bifurcation. In particular, the bifurcations were explained by
atomic double- and triple-well potentials in the rotating phase
space. Due to fluctuating atomic motions resulting from
spontaneous emissions, population transfer occurs between the two
states of dynamic double well, which tends to equalize the
population of each state [Fig. \ref{fig:1Image}(a)]. This atomic
transition between the two states oscillating in position space
(Fig. 1) was confirmed by observing the temporal recovery of the
population symmetry after emptying one state. In this case, the
recovering rates are equivalent to the transition rates
\cite{cnat2}.

It is interesting to observe that the population symmetry, which
is equivalent to zero spontaneous magnetization in the Ising spin
system, is only maintained below a certain critical value of the
total atomic number. Above the critical number, however, we have
observed SSB of atomic population, as shown in Fig.
\ref{fig:1Image}(b). The SSB can be observed under wide
experimental conditions of modulation frequency $f$ and amplitude
$h$, from super-critical to sub-critical bifurcation regions. The
atomic populations were simultaneously measured by resonant
absorption of a weak probe laser. The typical experimental
parameters are as follows: magnetic-field gradient along the
atomic oscillation direction ($z$-direction) $b$ = 14 G/cm,
cooling laser detuning $\delta$ = -2.6 $\Gamma$, and laser
intensity in the $z$-axis $I_z$ = 0.039 $I_s$ ($I_s$ is the
averaged saturation intensity, 3.78 mW/cm$^2$). The intensity on
the transverse axes is typically 5 times larger than that of the
$z$-axis. The measured trap frequency is 43.6 ($\pm$ 2.4) Hz
whereas the damping coefficient is 160.4 ($\pm$ 33) s$^{-1}$,
which is about three times larger than that expected in the
Doppler theory \cite{cnat3}.

\begin{figure}[ht]
\centering
\includegraphics[scale=0.43]{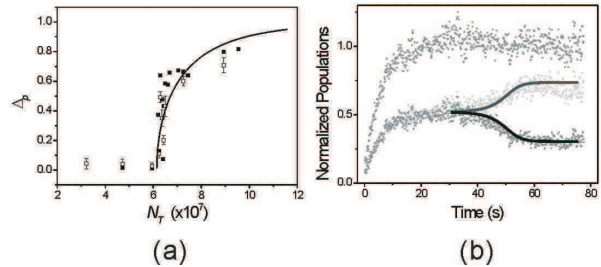}
\caption{(a) Experimental data of the normalized population
difference $\Delta_p$ versus the total atomic number $N_T$. Here
$f = 88$ Hz (= 2.0$f_0$) and $h$ = 0.86. The solid curve is the
fitting by the Ising model function with $N_c =6.1 (\pm 0.4)\times
10^{7}$. (b) Temporal evolution of SSB. The split theoretical
curves represent the normalized populations of the two dynamic
states ($N_T =6.5 \times 10^7$ and $\Delta_p =0.52$). The simple
sum of each population, i.e., the normalized total atomic number,
is also shown on the top.} \label{fig:2Measurement}
\end{figure}

Figure \ref{fig:2Measurement}(a) presents the normalized
population difference between the two dynamic states (1 and 2
with $N_1 > N_2$), $\Delta_p$ = ($N_1$ - $N_2$)/$N_T$ versus the
total atomic number, $N_T$ = $N_1$ + $N_2$. As shown in the
figure, the main control parameter of SSB is $N_T$ so that SSB
(or the Ising-class phase transition) occurs above the critical
number $N_c$. We have measured $N_1$ and $N_2$ by two independent
methods: CCD images (filled black boxes) and probe absorption
(empty boxes). $N_T$ was varied by adjusting the intensity of the
repumping laser while all the other trap parameters remain fixed.
Note that we did not find any other control parameters other than
$N_T$: for instance, the intensity imbalance between the $+z$ and
$-z$ laser beams did not contribute to SSB for the imbalance of
up to 20$\%$, beyond which the atomic limit cycle motions were
not sustained.

We have measured the critical number $N_c$ at various experimental
parameters of $f$ and $h$. For example, at $h=0.86$, $N_c$
decreases gently from $7.9 \times 10^{7}$ to $4.1 \times 10^{7}$
as $f$ increases from $1.95 f_0$ to $2.4 f_0$ ($f_0$ is the MOT
trap frequency along the $z$-axis). At $f=2.1 f_0 \; (= 90$ Hz),
$N_c$ also decreases from $11.9 \times 10^{7}$ to $6.2 \times
10^{7}$ as $h$ increases from 0.64 to 0.86. In brief, $N_c$
increases with the transition rate $W$: when $f$ or $h$
increases, $W$ decreases (see Ref. \cite{cnat2} for details), and
consequently $N_c$ becomes decreased. However, SSB is not observed
outside the above region of parameters, that is, near the
super-critical or sub-critical bifurcation points. Around the
super-critical bifurcation point, $W$ becomes too large to load
atoms enough to produce SSB in our experimental system. Near the
sub-critical bifurcation point, on the other hand, despite the
low $N_c$, only the population of the central stationary state
among the triple wells increases, whereas those of the two dynamic
states do not increase above $N_c$ \cite{cnat2}.

In order to have a better understanding of SSB, we have
investigated the temporal evolution of the populations of each
state when SSB occurs. Figure \ref{fig:2Measurement}(b) shows the
atomic populations recorded by the absorption of the probe laser.
As can be found, in the initial loading stage, the number of
atoms in each state increases at the same rate and their growth
is indistinguishable with each other. As the loading process is
finished at about 20 s elapse, however, the population of one
state increases whereas the other state is depopulated. The fact
that the total atomic number is conserved within experimental
errors during the SSB process indicates that SSB originates not
from different loading rates to each state but from the transfer
of atoms from one state to the other. Moreover, when we place a
kicking laser near the center of the two dynamic states in order
to block any transitions between the states, the population
symmetry is recovered. These evidences confirm that SSB occurs
due to the atomic transfer between the two states.

%\section{Intuitive explanations}
Based on the fact that SSB appears above the critical number, one
may conjecture that the underlying mechanism of SSB is related to
the collective effects of atoms occurring between the two dynamic
states \cite{Dalibard, Wieman, Bagnato, Wilkowski}. There are two
such collective mechanisms associated with the MOT atoms. One is
the shadow effect caused by absorption of the cooling lasers in
the $z$-axis due to atoms in one of the two states, which results
in the reduction of the laser intensity for atoms in the other
state. The other is the reradiation effect that arises when an
atom reabsorbs photons that are spontaneously emitted by another
atom, which produces the repulsive Coulomb-like forces between
the two atoms.

The reradiation effect, in fact, contributes as an obstacle to
SSB. As the number of atoms in one of the two states becomes
dominated due to fluctuations, the repulsive reradiation force
becomes bigger outside the more-populated atomic cloud. As a
result, this effect prevents atoms in the smaller-number state
from being transferred to the larger-number state, which results
in the recovery of population symmetry between the two states. On
the other hand, the shadow effect accelerates atoms to move from
the smaller-number state to the larger-number state: due to the
bigger shadow effects associated with the larger-number state, the
net atomic force is directed toward the larger-number state, which
enhances SSB further.

%\section{Calculations and Simulations}
For a theoretical understanding of SSB process, we have adopted
the phase-space Hamiltonian-function formalism developed in Ref.
\cite{Dykman} to account for the transitions between the dynamic
double wells \cite{cnat2}. We then have generalized the approach
to include the shadow effect as well as the reradiation force.
This approach provides quantitative analysis of nearly all the
fundamental characteristics of SSB such as the critical number and
the temporal evolution.

\begin{figure}[ht] \centering
\includegraphics[scale=0.38]{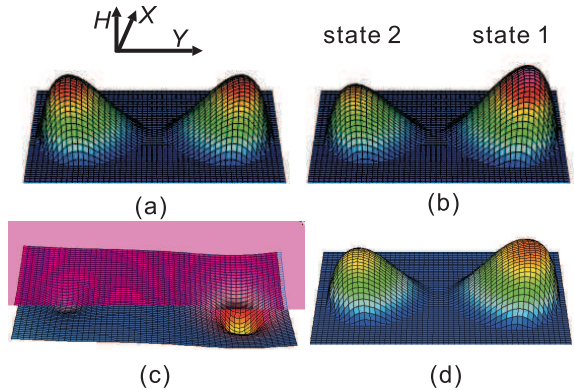}
\caption{{(a) Hamiltonian function $H^0_i(X_i,Y_i)$ [Eq.
(\ref{H_i})] for the symmetric state without collective effects.
(b) $H^0_i + H^S_{i}$ with the shadow effect included, which
leads to SSB in the phase-space potential. (c) Reradiation
interaction $H^R_{i}$, which opposes SSB. (d) $H^0_i + H^S_{i} +
H^R_{i}$, which shows slightly reduced SSB with respect to (b).
Here $N_1 = 3 \times 10^8$, $N_2 = 1 \times 10^8$, $I_z = 0.035
I_s$, $I_T$ = 8$I_z$, $b = 14$ G/cm, $\delta = -2.5 \Gamma$, $f =
2f_0$, $h = 0.9$, and the transverse (longitudinal) spatial width
of the atomic cloud is $R_{\rho} = 1$ mm ($R_{z} = 2$ mm). }}
\label{fig:3model}
\end{figure}

From the Doppler equation of MOT \cite{cnat2, cnat1}, one can
derive the phase-space Hamiltonian function of an $i$-th atom
without the collective effects as,

\begin{eqnarray}
H^0_i(X_i,Y_i) =\frac{1}{2}(\mu -1)X_i^2 + \frac{1}{2}(\mu
+1)Y_i^2 - \frac{1}{4}(X_i^2 +Y_i^2)^2\, , \label{H_i}
\end{eqnarray}
where $\mu = {2 ( f - 2 f_0 )}/{h f_0}$, $X_i$ and $Y_i$ are the
two scaled canonical variables in the rotating phase space, as
represented in Fig. \ref{fig:3model}(a). The total Hamiltonian
$H_0$ without the interaction terms is just the summation of each
$H^0_i$, $H_0 = \sum_{i=1}^{N_T} H^0_i$. Note that, in our
analysis, the two oscillating atomic states can be approximated
as a static double well in the phase space [Fig.
\ref{fig:3model}(a)]. For example, the right- and left-side cloud
with respect to the center of the limit cycle motion along the
$z$-axis corresponds to the positive and negative state,
respectively, in the $Y$ axis at a given modulation phase of
0$^{\circ}$. Note also that the maximum points in the phase-space
potential in Fig. \ref{fig:3model} indicate attractors.

Let us first consider the shadow effect that is responsible for
SSB. At the specific modulation phase of 0$^{\circ}$, we assume
the right-side atomic cloud is the state 1 and the left-side
cloud is the state 2. Because of the Zeeman shift, atoms in state
2 and state 1 absorb preferentially the cooling lasers
propagating in the $+z$ and $-z$ direction, respectively. Now we
consider another $j$-th atom in state 1 or 2, which absorbs the
laser by the amount $I^j_{A}$ = $\sigma^j_{L} n_{\rho} I_z$,
where $\sigma^j_{L}$ is the absorption cross-section of the $j$-th
atom and $n_{\rho}$ is the density of atoms in the $xy$-plane. One
can then easily find that each laser photon absorbed by the $j$-th
atom effectively results in a cooperative force (or acceleration)
on the $i$-th atom concerned, whose magnitude is given by $C_S
\sigma^j_{L} n_{\rho} I_z/I_s$. That is, when the atom $j$ is in
state 1 (i.e., in the right-side cloud absorbing the photons
propagating in the -$z$ axis), the direction of the effective
force experienced by the $i$-th atom is positive along the
$z$-axis, whereas it is negative when the $j$-th atom is in state
2. If one considers, for convenience, the $i$-th atom is near the
center, the net effective force exerted on the $i$-th atom is
given by $\sum_{j} C_S I^j_{A}/I_s$ = ($N_1 - N_2$) $C_S
I_{A}/I_s$. Therefore the $i$-th atom is transferred to the
larger-number state on the right (state 1) [Fig.
\ref{fig:3model}(b)]. Note that if one considers the $\pi$ phase
of modulation, although the location of state 1 (2) is now
exchanged to the left (right), the net force is still directed to
the larger-number state of 1.

The Hamiltonian $H^S_{i}$ for the shadow effect is then derived
as, when summed over $j$ in the states 1 and 2,
\begin{eqnarray}
H^S_{i}(X_i,Y_i) =& & \sum_{j \in {1, 2}}
(\alpha_j Y_i + H{'}_{j}) \, , \nonumber \\
=& & \alpha (N_1 - N_2) Y_i + (N_1 + N_2) H{'} \, , \label{H Si}
\end{eqnarray}
where $\alpha_j = \alpha = 2 C_S \sigma_L I_z/I_{s} \pi^2 \beta
\zeta^{3/2} \eta f 2 \pi R_{\rho}^2$ ($\sigma^j_{L}$ is assumed
independent of $j$), $C_S=\hbar k \Gamma /2 m (1+ 4
\delta^2/\Gamma^2)$, $\zeta = 2 \pi h f_0^2/\beta f$, $\eta =
\sqrt{4 \pi \beta f f_0^2 /3 A_0 (\beta^2 + 4 \pi^2 f_0^2 )}$,
$\beta$ is the damping coefficient, and $A_0$ is the coefficient
of the third-order term in the Doppler equation of MOT
\cite{cnat1}. $H{'}$ is a given coefficient that is practically
independent of $j$, with no contribution to SSB. Here $\sigma_L$
is regarded as having no dependence on velocity and position of
atoms, which are assumed uniformly distributed in the $xy$-plane.
We will discuss later about more realistic treatment of the
shadow effect with Monte-Carlo simulations. As shown in Fig.
\ref{fig:3model}(b), the shadow effect makes the potential of the
larger-number state (state 1) deeper, whereas that of the
smaller-number state (state 2) shallower. As a result, more atoms
will be transferred from state 2 to state 1, resulting in SSB of
the atomic population. In fact, however, there are competitions
between the shadow-induced SSB and the fluctuation-induced
symmetry-recovering transition. Therefore the critical number is
determined by the balance between the shadow effect and the
diffusion.

Let us now consider the symmetry-preserving reradiation effect.
Figure \ref{fig:3model}(c) presents the results of reradiation
interaction $H^R_{i}$, which reduces the SSB effect (detailed
expression of $H^R_{i}$ will be given elsewhere). Briefly
speaking, the reradiation effect increases the critical number by
reducing $\alpha$ to $\alpha - C_{R}$, where $C_R = I_T
\sigma_L^2/4 \pi c [3 \pi +8 (4 + 3 \ln2)]/6 \pi^2 R_z \beta
\zeta^2 \eta^2 f$. The calculation also shows that, if the
transverse laser intensity is over 10 times larger than that of
the $z$-direction laser, the reradiation effect dominates over
the shadow effect, which inhibits SSB for every $N_T$. In
practice, we have experimentally observed that the recovery of
symmetry appears at about 20 times the $z$-laser intensity, which
is a strong and independent evidence that the reradiation hinders
SSB.

\begin{figure}[ht]
\centering
\includegraphics[scale=0.43]{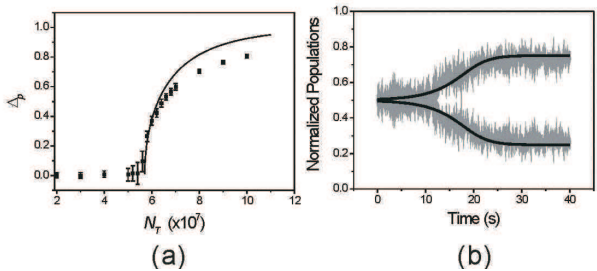}
\caption{{(a) $\Delta_p$ vs $N_T$, obtained by Monte-Carlo
simulations with $10^3$ atoms. The solid curve is a plot of Eq.
($\ref{Ising}$). (b) Simulation curves for the temporal evolution
of SSB, in good agreement with the experimental data (Fig.
\ref{fig:2Measurement}(b)). Here $W_0$ = 1.0 s$^{-1}$ and $N_T$ =
1.1$N_c$. }} \label{fig:4Compare}
\end{figure}

Let us discuss the temporal evolution of SSB, which can be
described by the simple rate equation, $d \Delta_p/dt =
W_{12}(1+\Delta_p )+W_{21}(1-\Delta_p)$, where the transition rate
$W_{12(21)}$ from state 1 (2) to state 2 (1) is $W_0 \exp(\mp
\Delta_p N_T/N_c)$ and $N_c = D/2 \alpha \zeta\
\sqrt{\mu+1}f(\mu)$. Here $W_0$ is the atomic transition rate
without the collective effects, $D$ is the phase-space diffusion
constant \cite{cnat2}, and $f(\mu)$ is an $O(1)$ function
\cite{Dykman}. Interestingly, the above rate equation leads to the
steady-state solution given by
\begin{eqnarray}
\Delta_p = \tanh(\Delta_p N_T/N_c)  \, . \label{Ising}
\end{eqnarray}
This is a representative equation of Ising-class phase
transition, which is plotted in Figs. \ref{fig:2Measurement}(a)
and 4(a).

To manifest further the relation with the Ising model, let us
consider a simplified model where each atom $i$ either belongs to
state 1 ($Y_i = \sqrt{\mu+1}$) or to state 2 ($Y_i =
-\sqrt{\mu+1}$). The activation energy $S_i$ due to the shadow
effect of the $j$-th atom is $- 2 \zeta f(\mu) \alpha_j Y_i$. The
total interaction energy is thus $\sum_{i} \sum_{j}^{N_T}$$S_{i}$
= -($J$/2) $(N_1-N_2)^2$, where $J$ = $2 \alpha \zeta f(\mu)
\sqrt{\mu+1}$. The free energy of this model system is then $F =
- (J/2) (N_T \Delta_p)^2$ + $D N_T \{ [(1+\Delta_p)/2] \ln
[(1+\Delta_p)/2] +[(1-\Delta_p)/2] \ln [(1-\Delta_p)/2] \} $. The
equilibrium value of $\Delta_p$ is determined by the condition
$\partial F/\partial \Delta_p$ = 0, which results in $\Delta_p$ =
$\tanh(\Delta_p N_T J/D)$ that is exactly the same as Eq.
($\ref{Ising}$) with $N_c$ = $D/J$. The simple analytical values
of $N_c$ are in qualitative agreement with the experimental
results of Fig. \ref{fig:2Measurement}(a).

We also have performed Monte-Carlo simulations with more realistic
consideration of the shadow effect and the reradiation force: we
included the dependence of $\sigma_L$ on the position and
velocity, the transverse laser-intensity profile, and the random
forces due to spontaneous emissions. Figure \ref{fig:4Compare}(a)
shows $\Delta_p$ versus $N_T$, which is very similar to the
experimental results in Fig. \ref{fig:2Measurement}(a). Figure
\ref{fig:4Compare}(b) presents the simulation curves for the
temporal evolution of SSB. Here we have just included the shadow
effect and used a diffusion constant that is 2.5 times the value
derived from the simple Doppler theory. When the reradiation force
is included in the simulations, however, $N_c$ is slightly
increased and SSB does not occur if the transverse laser
intensity is over 10 times the $z$-laser intensity. In
conclusion, nonlinear dynamic study of driven cold atoms may be
useful for dynamic phase transition and temporal dependence of
critical phenomena.

%\section{Conclusions}

\acknowledgements This work was supported by the CRI Project of
the MOST of Korea. H.R.N. was supported by Korea Research
Foundation Grant (KRF-2004-041-C00149).

%\newpage
%\centerline{\bf FIGURES} \vskip.5in

%\newpage

\end{document}